# Call Admission Control based on Adaptive Bandwidth Allocation for Multi-Class Services in Wireless Networks

Mostafa Zaman Chowdhury[a], Yeong Min Jang[a], and Zygmunt J. Haas[b]
[a] Department of Electronics Engineering, Kookmin University, Korea
[b] *Wireless Networks Lab,* Cornell University, Ithaca, NY, 14853, U.S.A
E-mail: *mzceee@yahoo.com*, *yjang@kookmin.ac.kr*, *zhaas@cornell.edu*

*Abstract*— Due to the fact that Quality of Service (QoS) requirements are not as stringent for non-real-time traffic types, as opposed to real-time traffic, more calls can be accommodated by releasing some bandwidth from the existing non-real-time traffic calls. If the released bandwidth to accept a handover call is larger than to accept a new call, then the probability of dropping a call is smaller than the probability of blocking a call. In this paper we propose an efficient Call Admission Control (CAC) that relies on adaptive multi-level bandwidth-allocation scheme for non-real-time calls. The features of the scheme allow reduction of the call dropping probability along with the increase of the bandwidth utilization. The numerical results show that the proposed scheme is able to attain negligible handover call dropping probability without sacrificing bandwidth utilization.

## I. Introduction

The trend of future wireless communication system is the decreasing of cell size and increasing the user mobility. These facts result the frequent handover in the wireless communication system. Whenever a session starts a call, the users always want to complete the session without any interruption. From the user's point of view, it is better to be blocked at the beginning than dropped at the middle of a call. Therefore, the mechanism to reduce the handover call dropping probability (HCDP) became a hot issue for the researchers in the field of wireless communication. Until now many researchers proposed their schemes to give higher priority for handover calls over new calls [1]. Most of the proposed schemes are based on the guard band. However, the guard band always reduces the bandwidth utilization.

There are diverse of traffic are found in wireless networks. They are classified in different categories [2]-[5]. The non-real-time traffic services are bandwidth adaptive [6], [7] and normally, they do not need Quality of Service (QoS) guarantees. In a system, more calls can be accommodate by reducing the allocated bandwidth for the existing non-real-time traffic calls and by reducing the requested bandwidth for the oncoming non-real- time traffic calls. However, the reduction of same amount of bandwidth from the non-real-time traffic calls to accept a handover call and a new call cannot reduce the HCDP significantly even though it reduces the call blocking probability. A multi-level bandwidth allocation for the non-real-time traffic calls proposed in this paper results negligible HCDP without reducing the resource utilization. Our proposed system can accept more handover calls over new calls. Also the minimum required bandwidth to accept a non-real-time handover call is less than that of a non-real-time new call in the proposed scheme. Consequently, the proposed scheme can accept more handover calls. Even though the proposed scheme blocks more new calls, the bandwidth utilization is not reduced. Compared to the adaptive bandwidth scheme, hard QoS scheme needs absolute reservation of network resources for specific traffic. Hard QoS scheme without guard channel cannot reduce the HCDP effectively. However, the guard channels in the hard QoS scheme increases the new call blocking probability and also reduce a measurable amount of bandwidth utilization.

The rest of this paper is organized as follows. Section II shows the system model for the proposed scheme. New call blocking probability and handover call dropping probability using the queuing analysis for the proposed scheme are shown in Section III. In Section IV, the numerical results for our proposed schemes are shown. Finally, conclusions are drawn in the last section.

## II. System Model for the Adaptive Bandwidth Allocation

Contemporary and future wireless network are required to serve different traffic types, which are classified by standardization bodies. Some of them required guaranteed bit rate (GBR) and some applications are not required guaranteed bit rate. The QoS parameters of the different traffic can be significantly different [2]-[5]. Normally the real-time services are categorized as GBR applications and non-real-time services are categorized as non guaranteed bit rate (NGBR) applications. Thus, under heavy traffic condition, the QoS of non-real-time services can be purposely degraded (e.g., by restricting bandwidth allocations), so that the QoS of real-time services is preserved (e.g., by maintaining low probability of blocking new calls or low probability of dropping handover calls).

The states of bandwidth degradation of traffic class $m$ is characterized by the bandwidth degradation factors $\gamma_m$, $\gamma_{m,n}$, and $\gamma_{m,n}$. Fig. 1 shows the multi-level bandwidth degradation model for the non-real-time applications of traffic class $m$.

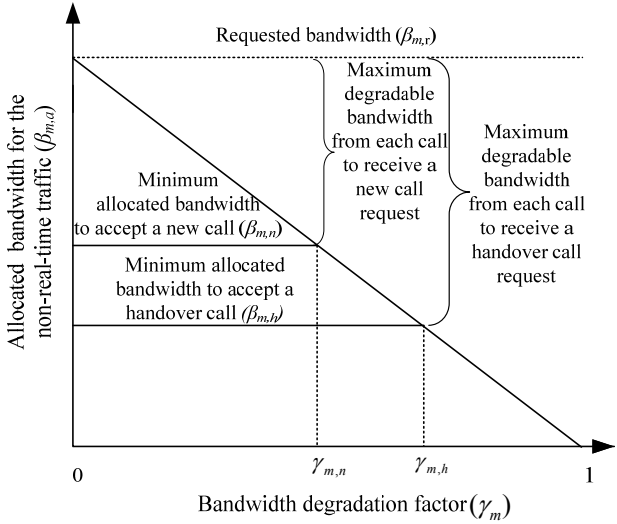

**Fig. 1**. Proposed multi-level bandwidth degradation model for non-real-time traffic services of class *m*

The bandwidth allocations $\beta_{m,a}$, $\beta_{m,n}$, and $\beta_{m,h}$ represent, respectively: the allocated bandwidth per call of already admitted calls of traffic of class *m*, the minimum allocated bandwidth per call to accept a new call of traffic of class *m*, and the minimum allocated bandwidth per call to accept a handover call of traffic of class *m*. Since the real-time traffic classes cannot degrade their bandwidth at all, the bandwidth degradation factor of all the real-time traffic classes equals zero. However, the system can release bandwidth from the existing non-real-time traffic calls (i.e., degrade the QoS of the non-real-time calls) to accept non-real-time and real-time traffic calls. The level of bandwidth degradation to accept a new call and a handover call are not, necessarily, equal. The fraction of the bandwidth that has been already degraded by an existing non-real-time call of traffic class *m*, the maximum fraction of bandwidth that can be degraded by an existing non-real-time call of traffic class *m* to accept a handover call request, and the maximum fraction of bandwidth that can be degraded by an existing non-real-time call of traffic class *m* to accept a new call request respectively are represented by the bandwidth degradation factor $\gamma_m$, $\gamma_{m,n}$, and $\gamma_{m,n}$ respectively.

In the system, new call arrival rate ($\lambda_n$), handover call arrival rate ($\lambda_h$) and average channel release rate ($\mu_c$) are shown using Fig. 2. $P_B$ and $P_D$ represent the original new call blocking probability and handover call dropping probability. The call arriving processes are assumed to be Passion. A new call that arrives in the system may either complete within the original cell or may handover to another cell or cells before completion. The probability of a call handover depends on two factors, (i) cell dwell time ($1/\eta$) and (ii) average call duration ($1/\mu$). However, the average duration of some non-real-time calls (e.g., file download) depends on the amount of allocated bandwidth. Another parameter, channel holding time ($1/\mu_c$) or average channel release rate ($\mu_c$), also depends on the two parameters (i) and (ii) above. In the traditional CAC, the average channel release rate ($\mu_c$) can be considered constant. However, in the adaptive bandwidth-allocation scheme, the average channel release rate ($\mu_c$) is decreased due to the increased call duration of some non-real-time calls. Thus, in our analysis, we assume variable channel release rate.

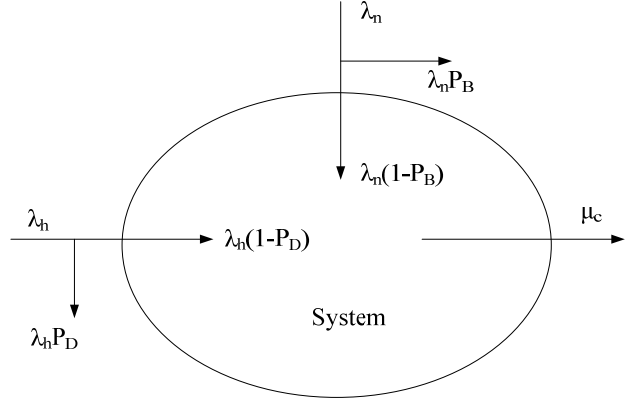

**Fig. 2.** The system scenario for the new call arrival rate, handover call arrival rate and average channel release rate

## III. Queuing Analysis

The proposed scheme can be modeled as an *M/M/K/K* queuing system (the value of *K* will be defined shortly). Suppose that the ratio of the calls arriving to the system for the *M* traffic classes is $a_1 : a_2 : \ldots : a_M$, where:

$$\sum_{m=1}^{M} a_m = 1 \qquad (1)$$

Suppose maximum number of calls that can be accommodated using the traditional hard QoS scheme is *N*. The Markov Chain for the proposed scheme is shown in Fig. 5. The maximum number of additional calls that can be supported by the proposed priority adaptive bandwidth allocation scheme is

$$S = \left\lfloor \frac{C \sum_{m=1}^{M} a_m \gamma_{m,h} \beta_{m,r}}{\sum_{m=1}^{M} \{a_m(1-\gamma_{m,h})\beta_{m,r}\} \sum_{m=1}^{M} \{a_m \beta_{m,r}\}} \right\rfloor \qquad (2)$$

The maximum number of calls that can be accommodated using the proposed adaptive bandwidth-allocation scheme is *K=(N+S)*. The maximal number of additional states of the Markov Chain in which the system accepts new call is

$$L = \left\lfloor \frac{C \sum_{m=1}^{M} a_m \gamma_{m,n} \beta_{m,r}}{\sum_{m=1}^{M} \{a_m(1-\gamma_{m,n})\beta_{m,r}\} \sum_{m=1}^{M} \{a_m \beta_{m,r}\}} \right\rfloor \qquad (3)$$

Due to apply of bandwidth degradation, the call duration of some of the non-real-time traffic calls are increased that causes the reduction of average service rate. The average channel release rate ($\mu_c$) for the proposed scheme is

$$\mu_c = \begin{cases} \mu_1 & for\ 0 < i \leq N \\ \mu_i & for\ N < i \leq N+S \end{cases} \qquad (4)$$

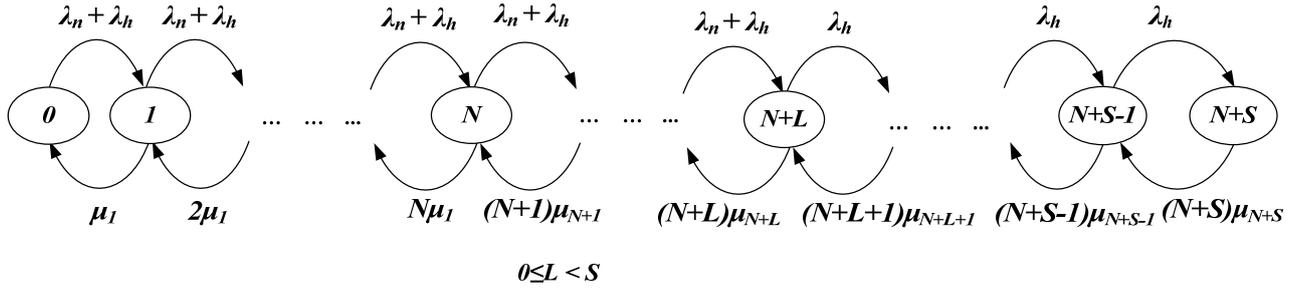

**Fig.3.** The Markov Chain of the proposed bandwidth-adaptive CAC

A new call in the proposed scheme is blocked if the state of the system calls is *(N+L)* or larger. However, a handover call is dropped if the state of the system calls is *(N+S)*. Thus, from (1) to (4), the call blocking probability ($P_B$) of an originating new call and the call dropping probability ($P_D$) of a handover call can be calculated as per equations (5) and (6) below.

$$P_B = \sum_{i=N+L}^{N+S} P(i) = (\lambda_n + \lambda_h)^{\left\lfloor \frac{C}{\sum_{m=1}^{M}\{a_m(1-\gamma_{m,n})\beta_{m,r}\}} \right\rfloor} \sum_{i=N+L}^{N+S} \frac{(\lambda_h)^{i-\left\lfloor \frac{C}{\sum_{m=1}^{M}\{a_m(1-\gamma_{m,n})\beta_{m,r}\}} \right\rfloor}}{i!\,(\mu_1^N) \prod_{p=N+1}^{i} \mu_p} P(0) \qquad (5)$$

$$P_D = P(N+S) = \frac{(\lambda_n + \lambda_h)^{\left\lfloor \frac{C}{\sum_{m=1}^{M}\{a_m(1-\gamma_{m,n})\beta_{m,r}\}} \right\rfloor} \lambda_h^{S-L}}{(N+S)!\,(\mu_1^N) \prod_{p=N+1}^{N+S} \mu_p} P(0) \qquad (6)$$

where $P(0) = \left[ \sum_{i=0}^{N} \frac{(\lambda_n + \lambda_h)^i}{i!\,\mu_1^i} + \sum_{i=N+1}^{N+L} \frac{(\lambda_n + \lambda_h)^i}{i!\,(\mu_1^N) \prod_{p=N+1}^{i} \mu_p} \right.$

$\left. + \sum_{i=N+L+1}^{N+S} \frac{(\lambda_n + \lambda_h)^{\left\lfloor \frac{C}{\sum_{m=1}^{M}\{a_m(1-\gamma_{m,n})\beta_{m,r}\}} \right\rfloor} (\lambda_h)^{i-\left\lfloor \frac{C}{\sum_{m=1}^{M}\{a_m(1-\gamma_{m,n})\beta_{m,r}\}} \right\rfloor}}{i!\,(\mu_1^N) \prod_{p=N+L+1}^{i} \mu_p} \right]$

## IV. Numerical Results

The numerical analyses for the proposed scheme are performed in this section. Table 1 shows the basic assumptions for the analyses. The call arriving process is assumed to be Poisson. The average cell dwell time is found to be 240 sec [8].

Table 1 Basic assumptions for the numerical analyses

| Assumptions for the traffic classes | | | | |
|---|---|---|---|---|
| | Traffic class (m) | Requested bandwidth by each call $\beta_{m,r}$ | $\gamma_{m,n}$ | $\gamma_{m,h}$ |
| Real-time services | Conversational voice (m=1) | 25 kbps | 0 | 0 |
| | Conversational video (m=2) (Live streaming) | 128 kbps | 0 | 0 |
| | Real-time game gaming (m=3) | 56 kbps | 0 | 0 |
| Non-real-time services | Buffered streaming video (m=4) | 128 kbps | 0.2 | 0.6 |
| | Voice messaging (m=5) | 13 kbps | 0.2 | 0.3 |
| | Web-browsing (m=6) | 56 kbps | 0.2 | 0.5 |
| | Background (m=7) | 56 kbps | 0.5 | 0.9 |
| Assumptions for the traffic environment | | | | |
| Average duration when $\beta_{m,r}$ is allocated during whole call duration | | 120 sec | | |
| Average user's speed | | 7.5 km/hr | | |
| Cell radius | | 1 km | | |

Fig. 4 shows that the proposed bandwidth adaptive scheme can reduce the handover call dropping probability even less than 0.0005 for very high traffic condition. This HCDP is also less than the 5% guard band scheme. However, the hard QoS scheme without any guard band and non-prioritized bandwidth adaptive schemes causes very high call dropping probability. The bandwidth utilization of the proposed scheme is maximum which is also equal to the non-prioritized bandwidth adaptive scheme. The bandwidth utilization comparisons are shown in Fig. 5. As mentioned before, the guard band always reduces the resource utilization that is also shown in Fig. 6. Hence, the performance of our proposed scheme is better than other schemes.

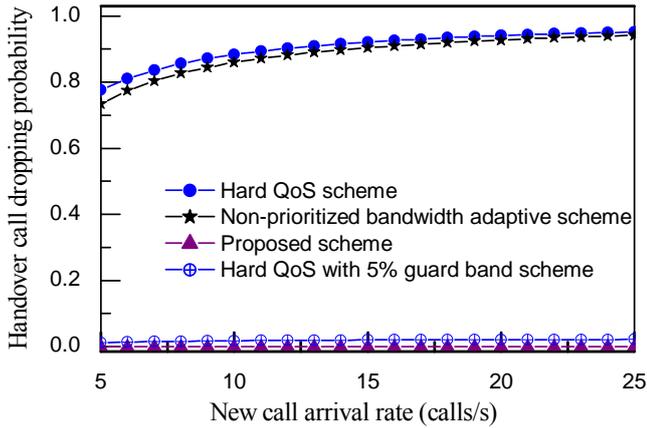

**Fig.4.** Comparison of handover call dropping probability during heavy traffic condition

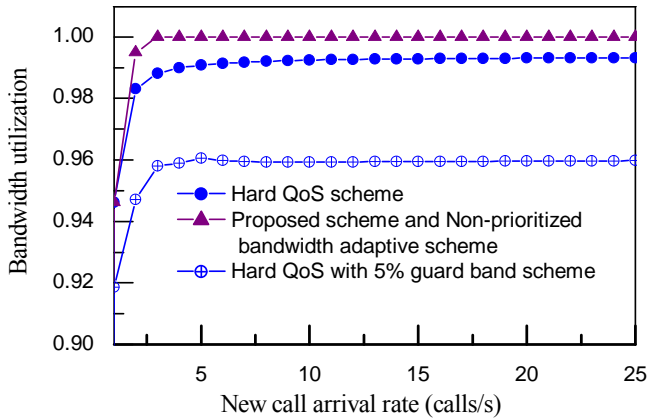

**Fig.5.** Comparison of bandwidth utilization

## V. Conclusions

We have shown that the proposed scheme is quite effective in reducing the HCDP without sacrificing the bandwidth utilization. While the proposed scheme blocks more new calls instead of dropping handover calls, the scheme also reduces the number of handovers and the average call duration as compared to the non-prioritized bandwidth-adaptive scheme. The proposed scheme is expected to be of considerable interest for future multi-service wireless networks, as the number of new traffic types with different QoS requirements is expected to further increase with the introduction of new applications.

## Acknowledgement

This work was supported by the IT R&D program of MKE/KEIT [10035362, Development of Home Network Technology based on LED-ID].